\documentclass[journal]{IEEEtran}

\usepackage[latin1]{inputenc}
\usepackage{times}
\usepackage{amsmath}
\usepackage{amssymb}
\usepackage{amsthm}
\usepackage{mathrsfs}
\usepackage{subcaption}
\usepackage{graphicx}
\usepackage{enumerate}
\usepackage[compress]{cite}
\usepackage{float}
\usepackage[acronym]{glossaries}
\usepackage{enumitem}
\usepackage{algorithmic}
\usepackage{algorithm}
\usepackage{multirow}
\usepackage{etoolbox}
\usepackage{subcaption}
\usepackage{booktabs}
\usepackage{lipsum}  
\usepackage{todonotes}

\AtBeginEnvironment{algorithmic}{\fontsize{6}{7}\selectfont}%{\small}
\newcommand{\bs}{\boldsymbol}

\DeclareMathOperator{\diag}{diag}

\DeclareMathOperator{\tr}{tr}

\newcommand{\vect}[1]{\bs{#1}}

\def\BibTeX{{\rm B\kern-.05em{\sc i\kern-.025em b}\kern-.08em
    T\kern-.1667em\lower.7ex\hbox{E}\kern-.125emX}}

\newtheorem{proposition}{Proposition}

\newacronym{3GPP}{3GPP}{third generation partnership project}
\newacronym{5G}{5G}{fifth generation}
\newacronym{6G}{6G}{sixth generation}
\newacronym{ASD}{ASD}{angular standard deviation}
\newacronym{AoA}{AoA}{angle of arrival}
\newacronym{AWGN}{AWGN}{additive white gaussian noise}
\newacronym{AP}{AP}{access point}
\newacronym{CF}{CF}{cell-free}
\newacronym{CF-M-MIMO}{CF-M-MIMO}{cell-free massive MIMO}
\newacronym{C2F-M-MIMO}{C$^2$F-M-MIMO}{clustered cell-free massive MIMO}
\newacronym{CB}{CB}{conjugate beamforming}
\newacronym{CPU}{CPU}{central processing unit}
\newacronym{CSI}{CSI}{channel state information}
\newacronym{DL}{DL}{downlink}
\newacronym{DMMSE}{D-MMSE}{distributed MMSE}
\newacronym{GMR}{GMR}{generalized maximum ratio}
\newacronym{iid}{i.i.d.}{independent identically distributed}
\newacronym{LMMSE}{L-MMSE}{local MMSE}
\newacronym{PMMSE}{P-MMSE}{partial MMSE}
\newacronym{LS}{LS}{least-squares}
\newacronym{LSFD}{LSFD}{large-scale fading decoding}
\newacronym{MF}{MF}{matched filtering}
\newacronym{mMIMO}{mMIMO}{massive multiple-input multiple-output}
\newacronym{MMSE}{MMSE}{minimum mean square error}
\newacronym{MS}{MS}{mobile station}
\newacronym{MR}{MR}{maximum ratio}
\newacronym{NC}{NC}{network controller}
\newacronym{NCB}{NCB}{normalized conjugate beamforming}
\newacronym{OBE}{OBE}{optimal bilinear equalizer}
\newacronym{PM}{PM}{pilot matching}
\newacronym{RCMMSE}{RC-MMSE}{reduced-complexity MMSE}
\newacronym{SE}{SE}{spectral efficiency}
\newacronym{SOC}{SOC}{second order cone}
\newacronym{SNR}{SNR}{signal-to-noise ratio}
\newacronym{SPA}{SPA}{slow power allocation}
\newacronym{TDD}{TDD}{time division duplex}
\newacronym{RV}{RV}{random variable}
\newacronym{UatF}{UatF}{use-and-then-forget}
\newacronym{UE}{UE}{user equipment}
\newacronym{ULA}{ULA}{uniform linear array}

\newacronym{2D}{2D}{two-dimensional}
\newacronym{3D}{3D}{three-dimensional}
\newacronym{BS}{BS}{base station}
\newacronym{CAP}{CAP}{compress-after-precoding}
\newacronym{CBP}{CBP}{compress-before-precoding}
\newacronym{CDF}{CDF}{cumulative distribution function}
\newacronym{CCDF}{CCDF}{complementary cumulative distribution function}
\newacronym{CoMP}{CoMP}{coordinated multipoint}
\newacronym{C-RAN}{C-RAN}{cloud radio access network}

\newacronym{DC}{DC}{difference of convex}
\newacronym{DoF}{DoF}{degrees of freedom}
\newacronym{FD}{FD}{full-dimensional}
\newacronym{FDD}{FDD}{frequency division duplexing}
\newacronym{FD-MIMO}{FD-MIMO}{full-dimensional MIMO}
\newacronym{GOPA}{GOPA}{globally optimal power allocation}
\newacronym{ICIC}{ICIC}{inter-cell interference coordination}
\newacronym{LCAP}{LCAP}{layered \gls{CAP}}
\newacronym{LCBP}{LCBP}{layered \gls{CBP}}
\newacronym{LOS}{LOS}{line-of-sight}

\newacronym{NLOS}{NLOS}{non-line-of-sight}
\newacronym{NOPA}{NOPA}{normalized optimal power allocation}
\newacronym{QoS}{QoS}{quality of service}
\newacronym{RHS}{RHS}{right hand side}
\newacronym{RRH}{RRH}{remote radio head}
\newacronym{rms}{rms}{root mean square}
\newacronym{SINR}{SINR}{signal-to-interference-plus-noise ratio}
\newacronym{UC}{UC}{user-centric}
\newacronym{UL}{UL}{uplink}
\newacronym{URA}{URA}{uniform rectangular array}
\newacronym{ZF}{ZF}{zero-forcing}
\newacronym{RZF}{RZF}{regularized ZF}

\newacronym{mMTC}{mMTC}{massive Machine Type Communicantions}
\newacronym{URLLC}{URLLC}{Ultra-Reliable Low-Latency Communications}

\begin{document}

\title{Pilot Decontamination Processing in Cell-Free Massive MIMO}

\author{Alberto \'Alvarez Polegre,~\IEEEmembership{Student~Member,~IEEE}, Luca Sanguinetti,~\IEEEmembership{Senior~Member,~IEEE}, and Ana Garc\'ia Armada,~\IEEEmembership{Senior~Member,~IEEE}

\thanks{A. \'Alvarez Polegre and A. Garc\'ia Armada are with the Department of Signal Theory and Communications, Universidad Carlos III de Madrid, Avda. Universidad 30, 28911, Legan\'es (Madrid), Spain (email: aalvarez@tsc.uc3m.es, agarcia@tsc.uc3m.es).

L. Sanguinetti is with the Dipartimento di Ingegneria dell'Informazione, University of Pisa, 65122 Pisa, Italy (email: luca.sanguinetti@unipi.it).

This work has been supported by project IRENE-EARTH (PID2020-115323RB-C33 / AEI / 10.13039/501100011033). L. Sanguinetti was in part supported by the Italian Ministry of Education and Research in the framework of the CrossLab project.

}\vspace{-0.5cm}
}
% The paper headers
%\markboth{In preparation, version \today}{Alv\'arez Polegre \textit{et al.}: A Sub-Optimal Cell-Free Massive MIMO Approach for Pilot Contamination Constrained Scenarios}

\maketitle

\begin{abstract}
	This letter focuses on the pilot contamination problem in the uplink and downlink of cell-free massive multiple-input multiple-output networks with different degrees of cooperation between access points. The optimum minimum mean square error processing can take advantage of large-scale fading coefficients for canceling the interference of pilot-sharing user-equipments and thus achieves asymptotically unbounded capacity. However, it is computationally demanding and can only be implemented in a fully centralized network. Here, sub-optimal schemes are derived that provide unbounded capacity with 
%	much lower complexity
	linear-growing complexity and using only local channel estimates but global channel statistics. This makes them suited for both centralized and distributed networks. In this latter case, the best performance is achieved with a generalized maximum ratio combiner that maximizes a capacity bound based on channel statistics only. 
\end{abstract}

\begin{IEEEkeywords}
	Cell-free massive MIMO, pilot contamination, uplink combining, downlink precoding, reduced-complexity minimum mean square error, generalized maximum ratio.
%	\vspace{-0.4cm}
\end{IEEEkeywords}
\vspace{-0.5cm}
\section{Introduction}
The interference generated by \glspl{UE} that transmit the same pilot sequence for channel acquisition is known as pilot contamination~\cite{Marzetta10,Bjornson18_unlimited}. This interference not only reduces the estimation quality but also makes the channel estimates statistically dependent. This latter effect makes it particularly hard to mitigate the interference (known as coherent interference) between these \glspl{UE}. Although existing in most networks, this problem has a greater impact on those where the large number of \glspl{UE} requires a high pilot-reuse factor. This was the case of cellular \gls{mMIMO} networks~\cite{Marzetta10,Bjornson18_unlimited} and has been inherited by the new \gls{CF} \gls{mMIMO}, but to a larger extent due to the \gls{CF} implementation\cite{Ngo17,Nayebi17}. In the beyond 5G era, \gls{CF} \gls{mMIMO} is seen as a promising solution to provide uniform service for \gls{mMTC} and \gls{URLLC}. These two use cases can turn pilot contamination into a totally unmanageable difficulty due to the amount of active \glspl{UE}.

Signal processing strategies play a key role when dealing with pilot contamination. Particularly, the \gls{MMSE} scheme was shown to achieve optimal performance in \gls{CF} \gls{mMIMO} networks, in contrast to the classical \gls{MR} strategy, which is severely limited by coherent interference\cite{Bjornson20}. The main drawback of \gls{MMSE} processing is that it comes at the cost of requiring a tight synchronized and centralized implementation, and high computational resources when the number of \glspl{AP} grows large. Considering capacity-limited wired connections and scalability concerns, such an implementation may limit the benefits of the \gls{CF} architecture. As far as we are concerned, there are not any low-complexity processing schemes that can deal with pilot contamination available in the \gls{CF} \gls{mMIMO} literature.

%This negative effect was believed to be a constrained for the system performance in terms of \gls{SE} not so long ago~\cite{Marzetta10}. Recently, it has been shown that spatially correlated channels (which capture more accurately the reality of massive \gls{MIMO} networks compared to the uncorrelated model) ease the multi-user interference cancellation even under severe pilot contamination~\cite{Bjornson18_unlimited}. The \gls{CF} massive \gls{MIMO} system can be seen as a large single-cell massive \gls{MIMO} system whose antennas are subject to different channel gains\cite{Ngo17}. Though initial literature on this topic relied on a simple conjugate beamformer for the combining/precoding operation, namely \gls{MR}, it has been demonstrated that \gls{MMSE}-based schemes largely outperform this strategy\cite{Bjornson20}. Nevertheless, the optimal \gls{MMSE} approach  requires centralized processing and involves high computational complexity. These two aspects can be hard to overcome in particular scenarios.

Building on \cite{Sanguinetti18}, in this letter we introduce a reduced-complexity, though sub-optimal, \gls{MMSE}-based combining scheme that achieves unbounded capacity in high pilot-contaminated \gls{CF} \gls{mMIMO} networks (rather than a distributed \gls{mMIMO} network) under different degrees of cooperation. The scheme can be implemented either in a centralized or a distributed fashion since only local channel estimates are required, in addition to global channel statistics. An alternative, yet optimal, scheme is also proposed by using a more conservative capacity bound that depends solely on statistical channel knowledge and provides an achievable rate when \gls{CSI} is not available for final data detection. This is the case of a decentralized network in which the channels are estimated at the \glspl{AP} while data detection is performed at the \gls{CPU}. These combiners are built based on the linear independence experienced by the global covariance matrices and the asymptotic behavior for a large number of \glspl{AP}, though they are valid under a finite regime. Neither of these processing schemes are constrained by pilot contamination and both show 
%a large reduction of computational complexity
linear-growing complexity. Thanks to the \gls{UL}-\gls{DL} duality\cite{DBS21-book}, precoding schemes can be obtained for the \gls{DL} segment. These schemes are also valid if a user-centric approach is taken (e.g.,~\cite{Buzzi17,DBS21-book}). 

\subsubsection*{Notation} 
%We use $\cdot^T$, $\cdot^*$, and $\cdot^H$ to denote transpose, conjugate and conjugate transpose, respectively. Uppercase and lowercase bold symbols are used to denote matrices and vectors, respectively. A complex column vector $\bs x$ of size $A$ is expressed by $\bs x \in \mathbb{C}^A$. We use $\mathbb{E}\left\{\cdot\right\}$ to denote expectation and indicate with $\diag(\cdot)$ the main diagonal components of a matrix. A complex normal distributed random variable $x$  with $\mu$ mean and $\sigma^2$ variance is denoted as $x\sim\mathcal{CN}(\mu,\sigma^2)$. Lastly, $\asymp$ expresses asymptotic equality.
The superscripts $^{T}$, $^*$ and $^{H}$ denote transpose, conjugate, and Hermitian transpose, respectively. We use $\mathcal{CN}({\bf 0},{\bf R})$ to denote the circularly symmetric complex Gaussian distribution with zero mean and covariance matrix ${\bf R}$. The expected value of a random variable $x$ is denoted by $\mathbb{E}\{ x \}$, while $\diag$ denotes the main diagonal components of a matrix. We use $a_n \asymp b_n$ to denote $a_n -b_n \to_{n\to \infty}0$ almost surely for two (random) sequences $a_n$, $b_n$.
%\vspace{-3mm}
%\subsubsection*{Acknowledgements}
%This work has been supported by project IRENE-EARTH (PID2020-115323RB-C33 / AEI / 10.13039/501100011033).
\vspace{-0.5cm}
\section{Network Model}\label{sec:network_model}
%\vspace{-0.2cm}
We consider a \gls{CF} \gls{mMIMO} network with $M$ single-antenna \glspl{AP}. The \glspl{AP} serve jointly $K$ single-antenna \glspl{UE}, and are connected via fronthaul links to a \gls{CPU}. The standard \gls{TDD} protocol of cellular \gls{mMIMO} is used~\cite[Sec. 2.3.2]{DBS21-book}, where the $\tau_c$
available channel uses are employed for: $\tau_p$ for \gls{UL} training phase, $\tau_d$ for \gls{DL} payload transmission and $\tau_u$ for \gls{UL} payload transmission. Clearly, $\tau_c\geq\tau_p + \tau_d + \tau_u$. 

The channel between \gls{AP} $m$ and \gls{UE} $k$ is $h_{mk} \sim \mathcal C \mathcal N (0,\beta_{mk})$ where $\beta_{mk}$ is the large-scale fading. We assume that $h_{mk}$ and $ h_{m'k}$ are uncorrelated for $m\neq m'$ due to the different location of the \glspl{AP}. We define $\bs h_{k}=[h_{1k},\ldots,h_{Mk}]^T\in\mathbb{C}^{M}$, which follows a $\mathcal{CN}({\bs 0}_{M}, {\bs{R}}_{k})$ distribution, where ${\bs{R}}_{k} = \diag(\beta_{1k},\ldots,\beta_{Mk})$ is diagonal. The matrices $\{{\bs{R}}_{k}: k=1,\ldots,K\}$ are assumed to be known, but practical estimation methods can be found in the literature (e.g.,~\cite[Sec. IV]{Sanguinetti2020}).

%\subsection{Pilot Transmission and Channel Estimation}
The same pilot sequence $\bs{\varphi}\in \mathbb{C}^{\tau_p}$ with $\|\bs{\varphi}\|^2=1$ is used by all the \glspl{UE} during the \gls{UL} training phase. The channel coefficients $\{h_{mk}: k=1\ldots, K\}$ can  be estimated either directly at \gls{AP} $m$ or at the the \gls{CPU}. In the latter case, \gls{AP} $m$ sends the received pilot signal to the \gls{CPU} via the fronthaul link. Regardless of where the channel estimation is carried out, the \gls{MMSE} estimate of ${h}_{mk}$ is~\cite[Sec. 4.1]{DBS21-book}
\begin{equation}
	\widehat{h}_{mk} =  \frac{\beta_{mk}}{ \psi_{m}}{\left(\frac{1}{\sqrt{\rho_p}}\bs{\varphi}^H{\bs y}_m\right)} %= \frac{\beta_{mk}}{ \psi_{m}}\widehat{h}_{mk}^{\text{LS}}
	\label{eq:h_mk_hat}
\end{equation}
where ${\bs y}_m$ is the received pilot signal
\begin{equation}
	{\bs y}_m = \sqrt{\rho_p}\sum_{k=1}^K  h_{mk} \bs{\varphi} + {\bs n}_{m}
\end{equation}
and $\rho_p$ is the pilot \gls{SNR} while $\bs n_{m} \in \mathbb{C}^{\tau_p}$ is the (normalized) noise vector with independent elements distributed as~$\mathcal{CN}(0,1)$. Also, we call
\begin{equation}\label{eq:pilotinterference}
	{\psi}_{m}=\beta_{mk} + \hspace{-1.2cm}\underbrace{\sum_{i=1, i\ne k}^K\hspace{-0.2cm}\beta_{mi}}_{\text{Interference from pilot-sharing UEs}} \hspace{-1cm} + \frac{1}{{\rho_p}}.
\end{equation}
%and ${1}/{\sqrt{\rho_p}}\bs{\varphi}^H\bs y_{m}$ is the \gls{LS} estimate of ${h}_{mk}$.
The estimates and estimation errors $\widetilde {h}_{mk}= h_{mk} - \widehat{ h}_{mk}$ are independent random variables distributed as~$\widehat{h}_{mk} \sim \mathcal{CN}\left(0,\gamma_{mk} \right)$ with $\gamma_{mk}=\beta_{mk}^2/ \psi_{m}$ and~$\widetilde{h}_{mk} \sim \mathcal{CN}\left(0,\varepsilon_{mk} \right)$ with $\varepsilon_{mk}  = \beta_{mk} - \gamma_{mk}$.
%\begin{equation}
%%	\varsigma^2_{mk}  = \beta_{mk} \left( 1 -  \frac{\beta_{mk}}{ \psi_{m}}\right).
%	\varsigma^2_{mk}  = \beta_{mk} - \gamma_{mk}.
%	\label{eq:E_mk}
%\end{equation}
The interference generated by the pilot-sharing \glspl{UE} in~\eqref{eq:pilotinterference} is known as pilot contamination, e.g.,~\cite{Bjornson18_unlimited}. It reduces the estimation quality, and makes the estimates correlated with $\widehat{h}_{mi} = \frac{\beta_{mi}}{\beta_{mk}}\widehat{h}_{mk}$.
Both effects deteriorate performance but only the second one is responsible of the coherent interference (e.g.,~\cite{Sanguinetti2020}). To perform coherent processing at multiple \glspl{AP}, knowledge of ${\bs h}_{k}$ is necessary. This is obtained as $\widehat{\bs h}_{k}=[\widehat{h}_{1k},\ldots,\widehat{ h}_{Mk}]^T\in\mathbb{C}^{M}$ and is distributed as $\widehat{\bs h}_{k} \sim \mathcal{CN}\left(\bs 0_M, \bs \Gamma_k \right)$
%\begin{equation}
%	\widehat{\bs h}_{k} \sim \mathcal{CN}\left(\bs 0_M, \bs R_{k} \bs \Psi^{-1} \bs R_{k}  \right)
%\end{equation}
with $\bs \Gamma_k =\bs R_{k} \bs \Psi^{-1} \bs R_{k}$ and $\bs \Psi = \diag({\psi}_1,\ldots,{\psi}_M)$ leading to $\bs \Gamma_k = \diag(\gamma_{1k},\ldots,\gamma_{Mk})$.
\section{Uplink Signal Combining}\label{sec:UL_combining}
We address now the \gls{UL} combining for both centralized and distributed processing. The optimal \gls{MMSE} scheme is set as the upper benchmark that our proposals will be compared to.
In the centralized scenario, each \gls{AP} $m$ acts only as a relay that forwards its received signals to the \gls{CPU}, which performs both channel estimation and data detection. Assuming error-free fronthaul links,
the joint \gls{UL} data signal 
$\bs y^{\text{ul}}$
%$\bs y^{\text{ul}} =[y_1^{\text{ul}},\ldots,y_M^{\text{ul}}]^T \in \mathbb{C}^{M}$ 
available at the \gls{CPU}
%is 
%\begin{equation}
%	\bs y^{\text{ul}}  = \sqrt{\rho_u}\sum_{k=1}^{K} \bs h_{k} s_k^{\text{ul}} + \bs n^{\text{ul}}
%%	\bs y^{\text{ul}}=[y_1^{\text{ul}},\ldots,y_M^{\text{ul}}]^T
%	\label{eq:data_symbol_UL_CPU}
%\end{equation}
%which
is mathematically equivalent to the signal model of a single-cell system where the correlated fading $\bs h_{k}$ has a diagonal spatial covariance matrix. At the \gls{CPU}, 
%the estimate of \gls{UE} $k$ data signal $s_k^{\text{\,ul}}$ 
the \gls{UE} $k$ data signal estimate
is obtained as $\widehat{s}_k^{\text{\,ul}} = \bs v_{k}^H \bs y^{\text{ul}}$
%\begin{equation}
%	\widehat{s}_k^{\text{\,ul}} = \bs v_{k}^H \bs y^{\text{ul}} %= \sum_{m=1}^{M}\bs v_{mk}^H \bs y_m^{\text{ul}}  	\label{eq:data_symbol_UL}
%\end{equation}
where $\bs v_{k}\in\mathbb{C}^{M}$ is the centralized combiner. With \gls{MMSE} channel estimation, an achievable \gls{SE} of \gls{UE} $k$ is 
$\text{SE}_k^{\text{ul}} = \frac{\tau_u}{\tau_c}\mathbb{E}\left\{\text{log}_2\left( 1 + \text{SINR}_k^{\text{ul}} \right) \right\}$
%\begin{equation}
%	\text{SE}_k^{\text{ul}} = \frac{\tau_u}{\tau_c}\mathbb{E}\left\{\text{log}_2\left( 1 + \text{SINR}_k^{\text{ul}} \right) \right\}
%	\label{eq:SE_CSI}
%\end{equation}
where the expectation is with respect to channel realizations and the effective \gls{SINR} is~\cite[Sec. 5.1.1]{DBS21-book}
\begin{equation}
	\text{SINR}_k^{\text{ul}} = \frac{| \bs v_k^H \widehat{\bs h}_k | ^2}{\bs v_k^H\left(\sum\limits_{i=1,i\neq k}^{K} \widehat{\bs h}_{i}\widehat{\bs h}_{i}^H {+}   {\bs Z} \right) \bs v_k}
	\label{eq:SINR_CSI}
\end{equation}
with $\bs Z = \diag({z}_1,\ldots,{z}_M)$ being diagonal with elements 
%\textcolor{blue}{${z}_m = \sum_{i=1}^{K}\varsigma^2_{mi} {+} {1}/{\rho_u}$}.
\begin{equation}\label{eq:Z_m}
	{z}_m = \sum_{i=1}^{K}\varepsilon_{mi} + \frac{1}{\rho_u}.
\end{equation}
This \gls{SE} is valid for any combiner $\bs v_{k}$ but requires the channel estimates $\widehat {\bs h}_k$ and estimation errors $\widetilde {\bs h}_k=[\widetilde{h}_{1k},\ldots,\widetilde{h}_{Mk}]^T$ to be independent. This condition is satisfied with the \gls{MMSE} channel estimator. An alternative bound that can be applied along with any channel estimator is the so-called \gls{UatF}.\footnote{The name comes from the fact that the channel estimates are used for computing receive combining vectors and then effectively forgotten before signal detection takes place.} This yields
$\text{SE}_k^{\text{ul,UatF}} = \frac{\tau_u}{\tau_c}\text{log}_2(1 + \text{SINR}_k^{\text{ul,UatF}})$,
%\begin{equation}
%	\text{SE}_k^{\text{ul,UatF}} = \frac{\tau_u}{\tau_c}\text{log}_2\left(1 + \text{SINR}_k^{\text{ul,UatF}}\right)
%	\label{eq:SE_uatf}
%\end{equation}
where $\text{SINR}_k^{\text{ul,UatF}} $ is given by~\cite[Sec. 5.1.2]{DBS21-book}
%which can lead to closed-form expressions instead on relying on Monte-Carlo simulations\cite{Ngo17}. The \gls{SINR} term is computed as
%
%where $\text{SINR}_k^{\text{UatF}}$ is given by \eqref{eq:SINR_uatf_general} shown at the top of next page.
%\begin{figure*}[t]
\begin{equation}
%	\begin{split}
\frac{\left| \mathbb{E} \left\{ |\bs v_k^H \bs h_k| \right\}  \right|^2} { \sum\limits_{i=1}^{K} \mathbb{E} \left\{\left| \bs v_k^H \bs h_{i} \right|^2 \right\} {-} \left| \mathbb{E} \left\{| \bs v_k^H \bs h_k |\right\}  \right|^2 {+} \frac{1}{\rho_u }\mathbb{E} \left\{ \left\| \bs v_k\right\|^2  \right\}  }
%	\end{split}
	\label{eq:SINR_uatf}
\end{equation}
and expectations are with respect to the channel realizations. Intuitively, $\text{SE}_k^{\text{ul,UatF}}$ is smaller than $\text{SE}_k^{\text{ul}}$ since it relies on a simplified implementation in which the channel
estimates are not used at the \gls{CPU} for signal detection.

Regarding the distributed network case, the \gls{MMSE} channel estimates are computed locally at the \glspl{AP} and are used to obtain local estimates of \gls{UE} data. This approach is more suitable for those use cases where low-latency is crucial. Let $ v_{mk}\in \mathbb{C}$ denote the local coefficient that \gls{AP} $m$ uses for \gls{UE} $k$. Then, 
%its local estimate of $s_{mk}$
the \gls{UE} $k$ local data estimate
is $\widehat s_{mk}^{\text{\,ul}}= v_{mk}^* y_m^{\text{ul}}$, 
with $y_m^{\text{ul}}$ the \gls{UL} signal at \gls{AP} $m$.
Any coefficient $ v_{mk}$ can be adopted. Unlike a fully centralized network, however, \gls{AP} $m$ can only use its own local channel estimates 
%$\{\widehat {h}_{mk}: k=1,\ldots K\}$
for the design of $v_{mk}$. The local estimates $\{\widehat s_{mk}^{\text{\,ul}}: m=1,\ldots,M\}$ are then sent to the \gls{CPU} for final decoding. We assume that the final estimate $\widehat s_{k}^{\text{\,ul}}$ is computed by averaging the local estimates, i.e., $\widehat s_{k}^{\text{\,ul}} =\sum_{m=1}^{M} \widehat s_{mk}^{\text{\,ul}}$.
%\begin{equation}
%	\widehat s_{k}^{\text{\,ul}} =\sum_{m=1}^M \widehat s_{mk}^{\text{\,ul}}. 
%\end{equation}
In a distributed network, the \gls{CPU} does not have knowledge of channel estimates and thus only the statistics can be utilized to estimate $\widehat s_{k}^{\text{\,ul}}$. Since $\text{SE}_k^{\text{ul,UatF}}$ does not rely on channel estimates, it can be used to compute an achievable \gls{SE} in a distributed network \cite[Sec. 5.2.1]{DBS21-book}. 
%In this case, an achievable SE of UE $k$ can be computed by using~\eqref{eq:SE_uatf} with $\bs v_{k}=[ v_{1k}, \ldots,v_{Mk}]^T\in\mathbb{C}^{M}$ since it does not rely on channel state information.
\vspace{-0.3cm}
\subsection{MMSE Combining}\label{subsec:MMSE_combining}
As in cellular \gls{mMIMO}\cite{Bjornson20,Sanguinetti2020}, the \gls{SINR} in~\eqref{eq:SINR_CSI} is a generalized Rayleigh quotient with respect to $\bs v_{k}$ and thus is maximized by the \gls{MMSE} combining vector, i.e.
\begin{equation}
	{\bs v}_{k} = \check{\bs v}_{k}=\left( \sum_{i=1}^{K} \widehat{\bs h}_{i} \widehat{\bs h}_{i}^H + \bs Z  \right)^{-1}\!\!\!\! 
	\widehat{\bs h}_k.
%	{\bs v}_{k} = \check{\bs v}_{k}=\left( \sum_{i=1}^{K} \sum_{m=1}^{M}\widehat{ h}_{mi} \widehat{ h}_{mi}^* + \bs Z  \right)^{-1}\!\!\!\! 
%	\widehat{\bs h}_k.
	\label{eq:combiner_MMSE}
\end{equation}
%with $\bs c_k=[c_{1k},\ldots,c_{Mk}]$. 
%\textcolor{blue}{This scheme will serve as an upper benchmark in terms of \gls{SE}.} 
We will now analyze the asymptotic behavior of $\text{SE}_k^{\text{ul}}$ with the \gls{MMSE} as $M \to \infty$. We assume that the \emph{global} covariance matrices $\{\bs{R}_{k}: k=1,\ldots,K\}$  are asymptotically linearly independent, as analytically defined in~\cite{Bjornson18_unlimited}. This assumption is fairly realistic since each entry of $\bs R_{k}$ is subject to a different channel gain. This is a key difference between the \gls{CF} architecture and traditional single-cell \gls{mMIMO} systems that allows \gls{CF} systems to achieve unbounded capacity even with single-antenna \glspl{AP}. Under this condition, and building on~\cite[Th. 1]{Bjornson18_unlimited}, the next proposition follows.
%\begin{proposition}~\cite[Th. 1]{Bjornson18_unlimited}\label{theorem:MMSE}
%	If \gls{MMSE} combining is used and $\{\bs{R}_{k}: k=1,\ldots,K\}$ are asymptotically linearly independent, then $\text{SE}_k^{\text{ul}}$ increases unboundedly as $M\to \infty$.
%\end{proposition}
\begin{proposition}\label{theorem:MMSE}
	If \gls{MMSE} combining is used and $\{\bs{R}_{k}: k=1,\ldots,K\}$ are asymptotically linearly independent, then $\text{SE}_k^{\text{ul}}$ increases unboundedly as $M\to \infty$.
\end{proposition}
\begin{IEEEproof}
%By applying the matrix inversion lemma and exploiting the block-diagonal structure of ${\bf Z}$, $\gamma_{1}$ in \eqref{eq:gamma1_MMSE} can be rewritten as
%\begin{align}
%\frac{\gamma_{1}}{M} = \sum\limits_{n=1}^N\frac{1}{M}{(\hat{\bf h}_1^n)}^{\Htran}{{({\bf Z}^n)}^{-1}} \hat{\bf h}_1^n - \frac{\left|\sum\limits_{n=1}^N\frac{1}{M}{(\hat{\bf h}_2^n)}^{\Htran}{{({\bf Z}^n)}^{-1}} \hat{\bf h}_1^n\right|^2}{\frac{1}{M} + \sum\limits_{n=1}^N\frac{1}{M}{(\hat{\bf h}_2^n)}^{\Htran}{{({\bf Z}^n)}^{-1}} \hat{\bf h}_2^n}
%\end{align}
%by also multiplying and dividing each term by $M$. 
	It follows easily from~\cite[App. B]{Bjornson18_unlimited} since the centralized case is basically a single-cell \gls{mMIMO} network with diagonal spatial correlation matrices.
\end{IEEEproof}
\vspace{-0.1cm}
\subsection{Reduced-Complexity MMSE Combining}
Although the \gls{MMSE} combining vector $\check{\bs v}_{k}$ is optimal and achieves unbounded capacity (as in cellular \gls{mMIMO}), it requires to compute the $M\times M$ matrix inverse in~\eqref{eq:combiner_MMSE} in every coherence block, which may be too computationally demanding when the network size is large. Note this translates into a computational complexity that grows with $M^3$. To overcome this issue, we follow the same approach in~\cite{Sanguinetti18}, and consider the asymptotic regime.
However, in contrast to a distributed cellular \gls{mMIMO} network, we consider the asymptotic regime in \gls{CF} \gls{mMIMO} systems to be achieved by letting $M\to \infty$. This consideration invalidates the straightforward use of those derivations available in \cite{Sanguinetti18} in a \gls{CF} scenario. Nevertheless, as shown in the Appendix, we can propose an alternative combiner $\bs v_{k}=\bar{\bs v}_{k}=[\bar{v}_{1k},\ldots,\bar{v}_{Mk}]^T$ with $\bar{v}_{mk}$ given by
\begin{equation}
	\begin{split}
	\bar{v}_{mk}
%		\bar{v}_{mk}&=\frac{1}{z_m}\sum_{i=1}^{K}\underbrace{\left[ \left(\bs B+\frac{1}{M}\bs I_K \right)^{-1}\right]_{ki}}_{b_{ki}} \widehat{ h}_{mi}\\
%		\bar{v}_{mk}&=\frac{1}{z_m}\sum_{i=1}^{K}\left[ \left(\bs B+\frac{1}{M}\bs I_K \right)^{-1}\right]_{ki} \widehat{ h}_{mi}\\
		=\frac{1}{\sum\limits_{i=1}^{K}\varepsilon_{mi} {+} \frac{1}{\rho_u}}\sum_{i=1}^{K}b_{ki}\widehat{ h}_{mi},
	\end{split}
	\label{eq:combiner_RCMMSE}
\end{equation}
where $b_{ki}$ is the $(k,i)$th element of the matrix $\left(\bs B+\frac{1}{M}\bs I_K \right)^{-1}$. The entries of $\bs B$ are given by
%\begin{equation}
%	\begin{split}
%		\left[ \bs B\right]_{ki}
%		&= \frac{1}{M} {\rm tr}\left({\bs R}_k\bs\Psi^{-1}{\bs R}_{i}{{\bs Z}}^{-1}\right) = \frac{1}{M}\sum_{m=1}^M \frac{\beta_{mk}\beta_{mi}}{\psi_m z_m}\\
%		&= \frac{1}{M}\sum_{m=1}^M \frac{\beta_{mk}\beta_{mi}}{\left(\sum\limits_{i=1}^K \beta_{mi}  + \frac{1}{{\rho_p}}\right)\left(\sum\limits_{i=1}^{K}\varsigma^2_{mi} {+} \frac{1}{\rho_u}\right)}.
%	\end{split}
%	\label{eq:Beta}
%\end{equation}
\begin{equation}
	\left[ \bs B\right]_{ki}
	= \frac{1}{M}\sum_{m=1}^M \frac{\beta_{mk}\beta_{mi}}{\left(\sum\limits_{i=1}^K \beta_{mi}  + \frac{1}{{\rho_p}}\right)\left(\sum\limits_{i=1}^{K}\varepsilon_{mi} {+} \frac{1}{\rho_u}\right)}.
	\label{eq:Beta}
\end{equation}
Since $\check{\bs v}_{k}$ reduces to $\bar{\bs v}_{k}$  in the limiting regime $M\to\infty$, it follows that also $\bar{\bs v}_{k}$ achieves unbounded capacity in accordance to Proposition 1.\footnote{In practice, $M$ will not be infinite. Numerical results will show that \eqref{eq:combiner_RCMMSE} works well for practical numbers of \glspl{AP}.} Compared to $\check{\bs v}_{k}$, however, the computational complexity required by $\bar{\bs v}_{k}$ is much lower. To see this, we use $\widehat{h}_{mk}$ to obtain
\begin{equation}
	\bar{ v}_{mk}=\bar{\varsigma}_{mk} \left(\frac{1}{\sqrt{\rho_p}}\bs{\varphi}^H\bs y_{m}\right)
	\label{eq:combiner_RCMMSE-reduced}
\end{equation}
where
\begin{align}
	\bar{\varsigma}_{mk} =\frac{1}{\left(\sum\limits_{i=1}^K \beta_{mi}  + \frac{1}{{\rho_p}}\right)\left(\sum\limits_{i=1}^{K}\varepsilon_{mi} + \frac{1}{\rho_u}\right)}\sum_{i=1}^{K}b_{ki}\beta_{mi}.
\end{align}
Under the assumption that coefficients
$\{\bar{{\varsigma}}_{mk}: m =1,\ldots,M\}$
are available, from~\eqref{eq:combiner_RCMMSE} the complexity required for the computation of $\bar{\bs v}_{k} = [\bar{ v}_{1k},\ldots, \bar{ v}_{Mk}]^T$ scales linearly with $M$, rather than as $M^3$ as with \gls{MMSE} combining. Therefore, we refer to $\bar{\bs v}_{k}$ as \gls{RCMMSE}. If one considers a centralized \gls{CF} \gls{mMIMO} network with limited computational resources, the \gls{RCMMSE} combiner seems to be the most suitable approach available in the literature\footnote{Though not shown here due to the lack of space, the \gls{RCMMSE} combiner outperforms the proposed \gls{LMMSE} schemes in \cite{Bjornson20} under full pilot reuse.}.
Likewise, notice that, if the large-scale coefficients are locally available at \gls{AP} $m$ (which is a reasonable assumption since they change every several coherence intervals and can be broadcasted by the \gls{CPU}), $\bar{v}_{mk}$ can be locally computed. This implies that, unlike the \gls{MMSE} combiner $\check{\bs v}_{k}$, it can also be adopted in a distributed network. In this case, however, the \gls{UatF} bound must be used to compute an achievable \gls{SE} since signal detection takes place without channel knowledge in a distributed network. 
%\textcolor{blue}{Lastly, it worth mentioning that a scalable approach can also be followed for the \gls{RCMMSE} scheme by letting each \gls{AP} to serve only a subset of \glspl{UE} (similar to the \emph{partial} approaches shown in \cite{Bjornson20_scalable})}.

%As the \gls{MMSE} scheme, the \gls{RCMMSE} combiner is a centralized scheme which requires of global channel estimation and only $\tau_c MN$ complex scalars are to be sent through the fronthaul links. However, the computation of \eqref{eq:combiner_RCMMSE} just scales quadratically, in contrast to the optimal \gls{MMSE} in \eqref{eq:combiner_MMSE}. Under this approach, \gls{SE} can be estimated with the bound shown in \eqref{eq:SE_CSI}. In addition, the \gls{RCMMSE} is also not asymptotically limited by pilot contamination.

\vspace{-0.2cm}
\subsection{Generalized Maximum Ratio Combining}\label{subsec:GMR_combining}
The \gls{MMSE} combiner is obtained as a matrix transformation of $\widehat{\bs h}_k$. Inspired by this, we now assume that ${\bs v}_{k} = {\bs W}_{k} \widehat{\bs h}_k$
%\begin{equation}
%	{\bs v}_{k} = {\bs W}_{k} \widehat{\bs h}_k
%	\label{eq:combiner_GMR_general}
%\end{equation}
where ${\bs W}_{k}$ is an arbitrary matrix to be optimized. Unlike~\eqref{eq:combiner_MMSE}, we assume that this optimization can only be done on the basis of channel statistics (rather than of channel estimates). 
%In this way, ${\bs W}_{k}$ can be precomputed at the CPU, and the computation of ${\bs v}_{k}$ in~\eqref{eq:combiner_GMR_general} reduces to only one matrix-vector multiplication, which requires $M^2$ (rather than $M^2$) complex multiplications. This is substantially lower than the complexity of MMSE. Moreover, the computation of~\eqref{eq:combiner_GMR_general} only uses the channel estimates of the UEs in the own cell.
The question is how to optimally design ${\bs W}_{k}$ in order to not incur a significant \gls{SE} loss. Following~\cite{Sanguinetti18,Neumann18},
% and under the asymptotic regime
and considering again the asymptotic regime in \gls{CF} \gls{mMIMO}, 
we can use the \gls{SINR} in~\eqref{eq:SINR_uatf} from the \gls{UatF} bound to obtain: 
\begin{align}\label{eq:approximationOfGamma}
 \!\frac{\left| \tr ( \bs{W}_k^{H}\vect{R}_{k})\right|^2}{\!\!\!\!\sum\limits_{i=1}^K\left|\tr \big(\bs \Psi^{-1} \vect{R}_{k}\vect{W}_k^{H}\vect{R}_{i} \big)\right|^2 \!\!+\!  \tr \big(\vect{W}_k\bs \Gamma_k \vect{W}_k^{H}\vect{U}_{k} \big)}\!\!\!
\end{align}
with $\vect{U}_{k} = \sum_{i=1}^K \bs R_i + \frac{1}{\rho_u} \bs I_M$ being diagonal. The matrix $\vect{W}_k$ that maximizes \eqref{eq:approximationOfGamma} is given by (e.g.,~\cite[Sec. VII]{Sanguinetti2020})
\begin{equation}
	\begin{split}
		\vect{W}_k = \bs U_k^{-1}\left(\sum_{i=1}^{K}a_{ki}\bs R_{i}\right)\bs R_{k}^{-1}
	\end{split}
	\label{eq:combiner_GMR_matrix}
\end{equation}
where $a_{ki}$ is the $(k,i)$th element of the matrix $\left(\bs A+\frac{1}{M}\bs I_K \right)^{-1}$. The entries of $\bs A$ are given by
\begin{equation}
	\begin{split}
		\left[ \bs A\right]_{ki}
%		&=\frac{1}{M}\tr\left(\bs R_{k}\bs\Psi_{}^{-1}\bs R_{i} \bs\Psi^{-1} \right)=\frac{1}{M}\sum_{m=1}^M \frac{\beta_{mk}\beta_{mi}}{\psi_m^2}\\
%		&
		= \frac{1}{M}\sum_{m=1}^M \frac{\beta_{mk}\beta_{mi}}{\left(\sum\limits_{i=1}^K \beta_{mi}  + \frac{1}{{\rho_p}}\right)\left(\sum\limits_{i=1}^K \beta_{mi} + \frac{1}{\rho _u}\right)}.
	\end{split}
	\label{eq:Alpha}
\end{equation}
Plugging~\eqref{eq:combiner_GMR_matrix} into ${\bs v}_{k} = {\bs W}_{k} \widehat{\bs h}_k$ yields ${\bs v}_{k}=\tilde{{\bs v}}_{k} = [\tilde v_{1k},\ldots,\tilde v_{Mk}]^T$ with (thanks to the diagonal structure of matrices) 
\begin{equation}
	\begin{split}
		\tilde v_{mk}
%		&=\frac{1}{\psi_m}\sum_{i=1}^{K}\left[ \left(\bs A+\frac{1}{M}\bs I_K \right)^{-1}\right]_{ki} \widehat{ h}_{mi}\\
%		&
		=\frac{1}{\sum\limits_{i=1}^K \beta_{mi} + \frac{1}{\rho _u}}\sum_{i=1}^{K}a_{ki}\widehat{ h}_{mi}.
	\end{split}
	\label{eq:combiner_GMR}
\end{equation}
We refer to \eqref{eq:combiner_GMR} as \gls{GMR} combining. As for \gls{MMSE}, we can analyze its asymptotic behavior as $M \to \infty$. 
%Considering those findings made in
With~\cite{Sanguinetti18,Neumann18}, the next proposition follows.
\begin{proposition} \label{theorem:GMF}
	If \gls{GMR} combining is used and $\{\bs{R}_{k}: k=1,\ldots,K\}$ are asymptotically linearly independent, then $\text{SE}_k^{\text{ul,UatF}}$ increases unboundedly as $M\to \infty$.
\end{proposition}
This proposition shows that \gls{GMR} has the same scaling behavior of \gls{MMSE} and \gls{RCMMSE} as $M\to \infty$. As with \gls{RCMMSE}, this is achieved by using only the channel statistics. Moreover, \gls{GMR} has the same complexity of \gls{RCMMSE} as it follows by plugging $\widehat{h}_{mk}$ into~\eqref{eq:combiner_GMR} to obtain 
\begin{equation}
	\tilde v_{mk}=\tilde {\varsigma}_{mk} \left(\frac{1}{\sqrt{\rho_p}}\bs{\varphi}^H\bs y_{m}\right)
	\label{eq:combiner_GMR-reduced}
\end{equation}
where
\begin{align}
	\tilde {\varsigma}_{mk} =\frac{1}{\left(\sum\limits_{i=1}^K \beta_{mi}  + \frac{1}{{\rho_p}}\right)\left(\sum\limits_{i=1}^K \beta_{mi} + \frac{1}{\rho _u}\right)}\sum_{i=1}^{K}a_{ki}\beta_{mi}.
\end{align}
The key difference with respect to \gls{RCMMSE} in~\eqref{eq:combiner_RCMMSE} is the scaling factor $\tilde {\varsigma}_{mk}$, which does not depend on the variances $\{\varepsilon_{mi}: i=1,\ldots,K\}$ of the channel estimation errors but on the large-scale fading coefficients $\{\beta_{mi}: i=1,\ldots,K\}$. This is because~\eqref{eq:combiner_GMR} maximizes $\text{SE}_k^{\text{ul,UatF}}$, which does not take into account the imperfect \gls{CSI} in the decoding process. As \gls{RCMMSE}, \gls{GMR} can also be used in a decentralized network. Whatsmore, anytime a distributed scenario with global statistics knowledge is considered, \gls{GMR} should be used since it is the optimal scheme for \gls{MR}-based processing.

\begin{table}
	\caption{Simulation parameters}
	\label{table1}\centering
	\begin{tabular}{l l}
		\toprule[0.4mm]
		\textbf{Parameter} & \textbf{Value}\\
		Carrier frequency											& 2 GHz\\
%		System bandwidth											& 20 MHz\\
		Path loss exponent ($\alpha$) 								& 3.76\\
		Shadowing standard deviation ($\sigma_{\chi}$) 				& 10\\
		Noise figure at receivers									& 7 dB \\
		\gls{DL} transmission power									& 200 mW \\
		\gls{UL} transmission power									& 100 mW\\
%		\gls{AP} antenna height										& 15 m \\
%		\gls{UE} antenna height										& 1.5 m \\
		Channel coherence interval length ($\tau_c$)				& 200 samples \\
		Training phase interval length ($\tau_p$)					& 1 sample \\
		Payload phase interval length ($\tau_d=\tau_u$)				& $\left(\tau_c-\tau_p\right)/2$ samples \\
		\bottomrule[0.4mm]
	\end{tabular}\vspace{-0.5cm}
\end{table}

%\vspace{-0.5cm}
\section{Downlink Signal Precoding}\label{sec:downlink_precoding}
%We now provide similar alternatives for the \gls{DL} segment. By letting $\bs w_{mk}$ be the $N$ beamforming vector, the transmitted signal from each \gls{AP} to \glspl{MS} results in
%\begin{equation}
%	\bs x_m = \sqrt{\rho_d} \sum_{k=1}^{K} \sqrt{\eta_{mk}} \bs  w_{mk} s_k^{\text{dl}},
%	\label{eq:x_m}
%\end{equation}
%where $\rho_d$ denotes the \gls{DL} transmission power and $s_k^{\text{dl}}$ the intended data symbol for \gls{MS} $k$ satisfying $\mathbb{E}\left\{ |s_k^{\text{dl}}|^2\right\}=1$. Then, the estimated data symbol at reception is
%\begin{equation}
%	\widehat{s}_k^{\text{dl}} = \sum_{m=1}^{M}\bs h_{mk}^H\bs x_m + n_k
%	\label{eq:data_symbol_DL}
%\end{equation}
%where $n_k\sim\mathcal{CN}(0,1)$ accounts for the receiver thermal noise.
We now consider the \gls{DL} and design the precoding schemes for centralized and decentralized networks. As in the \gls{UL}, the most advanced \gls{DL} implementation is a fully centralized operation, where the \gls{AP} only act as relays that transmit signals generated by the \gls{CPU}. A distributed operation is also possible in the \gls{DL} where the \gls{CPU} encodes the \gls{DL} data signals $\{s_k^{\text{dl}}: k=1,\ldots,K\}$ and sends them to the \gls{AP}, which selects the precoding coefficients on the basis of local estimates. This is the key difference between the two operation modes in the \gls{DL}. The information available at the \gls{UE} for signal detection is the same with both implementations. Since there are no \gls{DL} pilots, in both cases \glspl{UE} have no knowledge of channel estimates and must rely on statistics. An achievable \gls{SE} can be computed for the two operation modes by using the channel hardening bound. This yields $\text{SE}_k^{\text{dl}} = \frac{\tau_d}{\tau_c}\text{log}_2(1 + \text{SINR}_k^{\text{dl}})$
%\begin{equation}
%	\text{SE}_k^{\text{dl}} = \frac{\tau_d}{\tau_c}\text{log}_2(1 + \text{SINR}_k^{\text{dl}})
%	\label{eq:SE_dl}
%\end{equation}
with \cite[Sec. 6.1.1]{DBS21-book}
\begin{equation}
	\text{SINR}_k^{\text{dl}} =
	\frac{\left| \mathbb{E} \left\{ |\bs h_k^H \bs  w_k |\right\}  \right|^2} {\sum\limits_{i=1}^{K} \mathbb{E} \left\{| \bs h_k^H \bs  w_{i}|^2 \right\} -  \left| \mathbb{E} \left\{| \bs h_k^H \bs  w_k| \right\}  \right|^2 + \frac{1}{\rho_d}  }.
	\label{eq:SINR_dl}
\end{equation}
As seen, $\text{SE}_k^{\text{dl}}$ depends on the precoding vectors $\{{\bs w}_i:i=1,\ldots,K\}$ of all \glspl{UE}. This stands in contrast to the \gls{UL} \gls{SE} and makes optimal precoding design hard. A common heuristic approach relies on the \gls{UL}-\gls{DL} duality~\cite[Sec. 6.1.2]{DBS21-book}, which holds between the \gls{UatF} bound and hardening bound. %The duality states that the SE achieved in the \gls{UL} can be achieved also in the \gls{DL}, if the \gls{UL} combining vectors are used as \gls{DL} precoding vectors and the \gls{DL} transmit power is allocated properly. 
Motivated by this duality, we select the \gls{DL} precoding vectors as $\bs w_{k} = \frac{\bs v_{k}}{\sqrt{\mathbb{E}\{| \bs v_{k}^H\bs v_{k}|\}}}.$
%
%
%We now consider the DL and propose precoding schemes, which are the counterparts of the combining schemes derived above.
%We can straightforwardly address the design of the precoding vectors thanks to the \gls{UL}-\gls{DL} duality, which can guarantee the same \gls{SINR} for both segments following a given power control coefficients set\cite[Sec. 6.1.2]{DBS21-book}. Then, an arbitrary precoding vector $\bs  w_{k}=[w_{1k},\ldots,w_{Mk}]^T$ will be computed as
%\begin{equation}
%	\bs w_{k} = \frac{\bs v_{k}} {\sqrt{\mathbb{E}\{| \bs v_{k}^H\bs v_{k}|\}}}.
%	\label{eq:UL-DL_duality}
%\end{equation}
Precoding vectors can be chosen on the basis of \gls{RCMMSE} or \gls{GMR} for either centralized or distributed networks. On the other hand, the \gls{MMSE} precoder can only be used with a  centralized implementation.
%If operated in a centralized fashion, the precoding vectors can be selected according to the \gls{MMSE} combiner in~\eqref{eq:combiner_MMSE}. On the other hand, they should be selected on the basis of \gls{RCMMSE} and \gls{GMR} with a distributed operation.
%Since there are not \gls{DL} pilots, \glspl{UE} have no knowledge of channel estimates and must rely on statistics instead. The \gls{SE} estimation must be computed by the channel hardening bound as \cite[Sec. 6.1.1]{DBS21-book}
%\begin{equation}
%	\text{SE}_k^{\text{dl}} = \frac{\tau_d}{\tau_c}\text{log}_2(1 + \text{SINR}_k^{\text{dl}}),
%	\label{eq:SE_dl}
%\end{equation}
%with
%\begin{equation}
%	\text{SINR}_k^{\text{dl}} =
%	\frac{\left| \mathbb{E} \left\{ |\bs h_k^H \bs  w_k |\right\}  \right|^2} {\sum\limits_{i=1}^{K} \mathbb{E} \left\{| \bs h_k^H \bs  w_{i}|^2 \right\} -  \left| \mathbb{E} \left\{| \bs h_k^H \bs  w_k| \right\}  \right|^2 + \frac{1}{\rho_d}  }.
%	\label{eq:SINR_dl}
%\end{equation}

\begin{figure}[t]
	\centering
	\begin{subfigure}[b]{0.45\textwidth}
		\includegraphics[width=1\columnwidth]{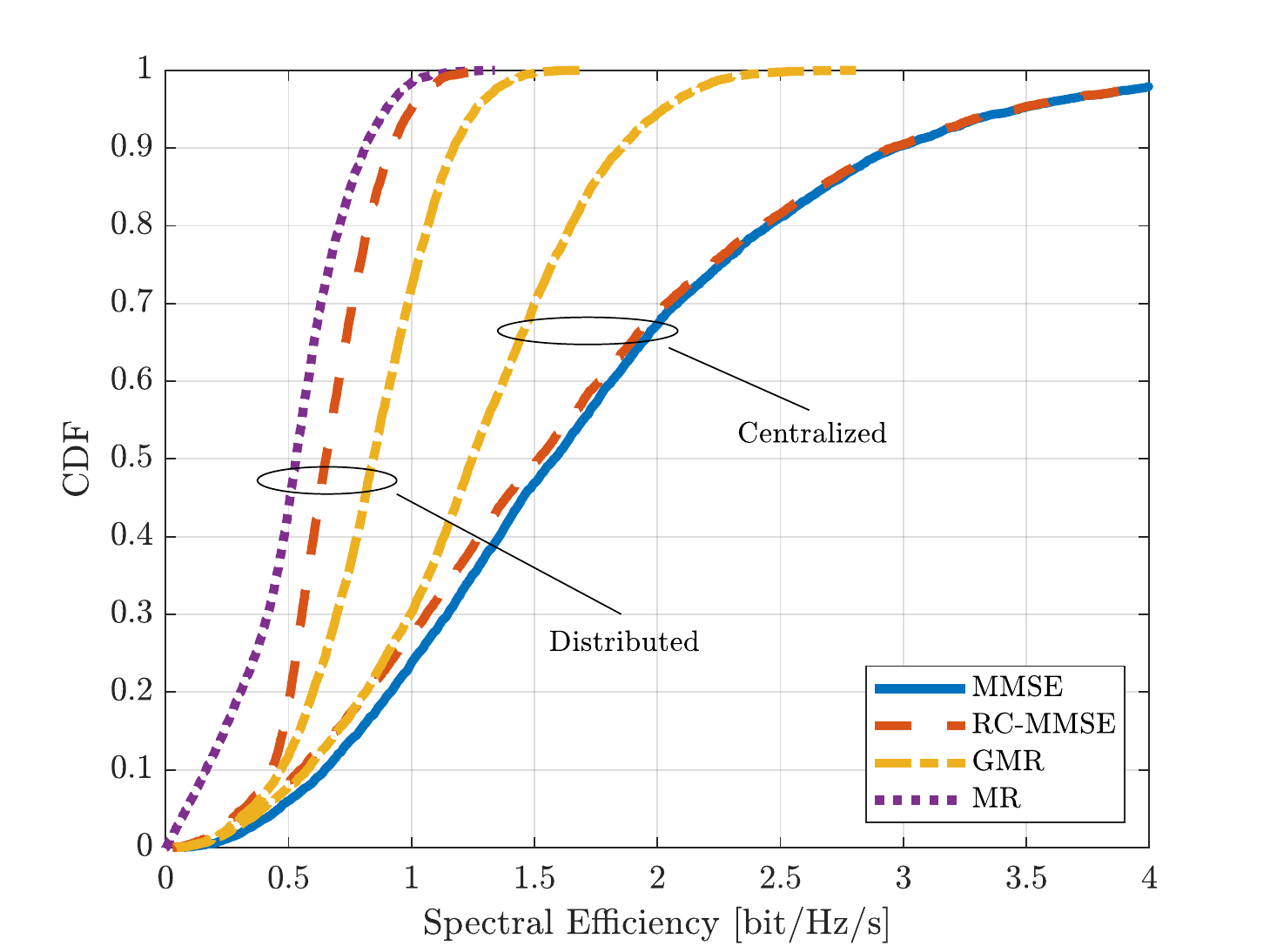}
		\caption{UL.}
		\label{fig:CDF_UL}
	\end{subfigure}
	\begin{subfigure}[b]{0.45\textwidth}
		\includegraphics[width=1\columnwidth]{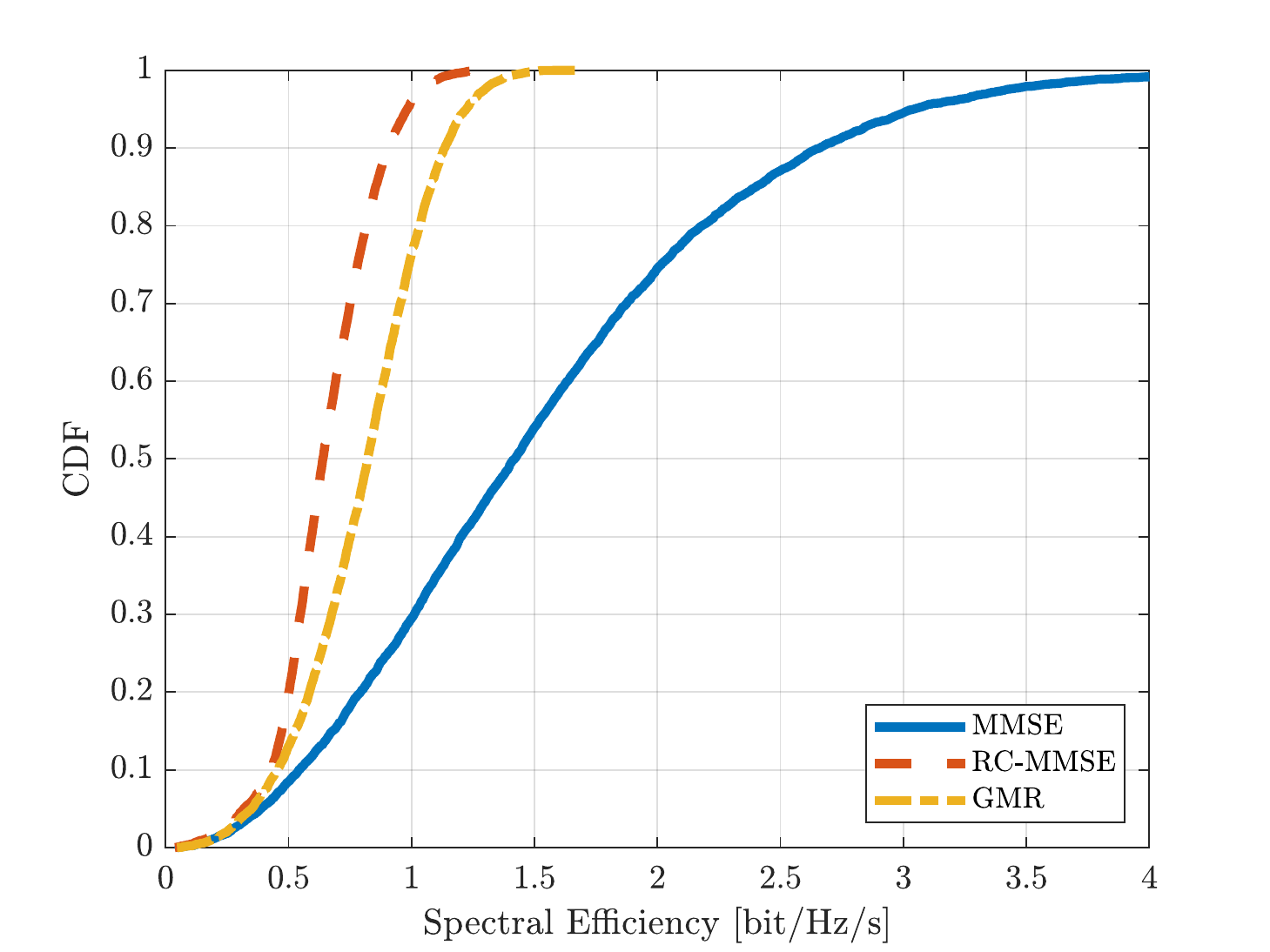}
		\caption{DL.}
		\label{fig:CDF_DL}
	\end{subfigure}
	\caption{CDF of average SE per UE for $K=20$.}\vspace{-0.4cm}
	\label{fig:CDF}
\end{figure}
%\vspace{-0.5cm}
\section{Performance Analysis}\label{sec:simulation_results}
Up next the system performance is shown for the proposed schemes. The simulation parameters are reported in Table \ref{table1}. \glspl{AP} will be randomly deployed within a square coverage area of side $D$. The large-scale fading coefficients are computed as
\begin{equation}
	\beta_{mk}[\text{dB}] = -35.3 + 10\alpha\text{log}_{10}(d_{mk}) + \chi_{mk} - \sigma_n^2
	\label{eq:L_mk}
\end{equation}
where $\alpha$ is the path loss exponent, $d_{mk}$ is the distance between \gls{AP} $m$ and \gls{UE} $k$, $\chi_{mk}\sim\mathcal{N}(0,\sigma_{\chi}^2)$ is the shadow fading component and $\sigma_n^2$ the noise variance.

The \gls{CDF} for the \gls{UL} average \gls{SE} per \gls{UE} is shown in Fig. \ref{fig:CDF_UL}. The number of \glspl{AP} is set to $M=100$ and $K=20$ randomly distributed \glspl{UE} are dropped with $D=250$ m. Both centralized and distributed versions of the \gls{RCMMSE} and \gls{GMR} combiners are considered. As upper and lower benchmarks, the centralized \gls{MMSE} and distributed \gls{MR} combiners are also presented. This latter scheme is computed by letting the entries of $\bs v_k$ be  $v_{mk}=\widehat{h}_{mk}$. Regarding centralized schemes, the proposed \gls{RCMMSE} scheme nearly matches the optimal combiner. This proves this new scheme is a very suitable alternative for lowering the required computational resources with little performance loss under severe pilot contamination. On the other side, the \gls{GMR} shows a considerable gap. We can notably point out that for low \gls{SNR} both \gls{RCMMSE} and \gls{GMR} match each other. This is due to the fact that the \gls{RCMMSE} combiner relies on having good channel estimates, while \gls{GMR} relies on channel coefficients instead. Focusing on the distributed approaches, the \gls{GMR} combiner outperforms the \gls{RCMMSE}. This is reasonable since \gls{GMR} is designed for maximization of the \gls{UatF} bound.

The \gls{DL} segment is shown in Fig. \ref{fig:CDF_DL} for the same setup. Power allocation coefficients with \gls{MMSE} and \gls{RCMMSE} are selected following a network-wide approach by letting $\eta_{mk}=\eta_k=1/K$, while these are selected as $\eta_{mk}=\sqrt{\beta_{mk}}/\sum_{i=1}^{K}\sqrt{\beta_{mi}}$
%\begin{equation}
%	\eta_{mk}=\frac{\sqrt{\beta_{mk}}}{\sum\limits_{i=1}^{K}\sqrt{\beta_{mi}}}
%	\label{eq:power_control_distributed}
%\end{equation}
for the distributed \gls{GMR} scheme. Thanks to the \gls{UL}-\gls{DL} duality, the channel hardening bound gives nearly the same \gls{DL} performance as the \gls{UatF} bound in \gls{UL}. This proves that \gls{RCMMSE} is unsuitable for a scenario in which only the statistics of channels are available.

\begin{figure}[t]
	\centering
	\includegraphics[width=.9\linewidth]{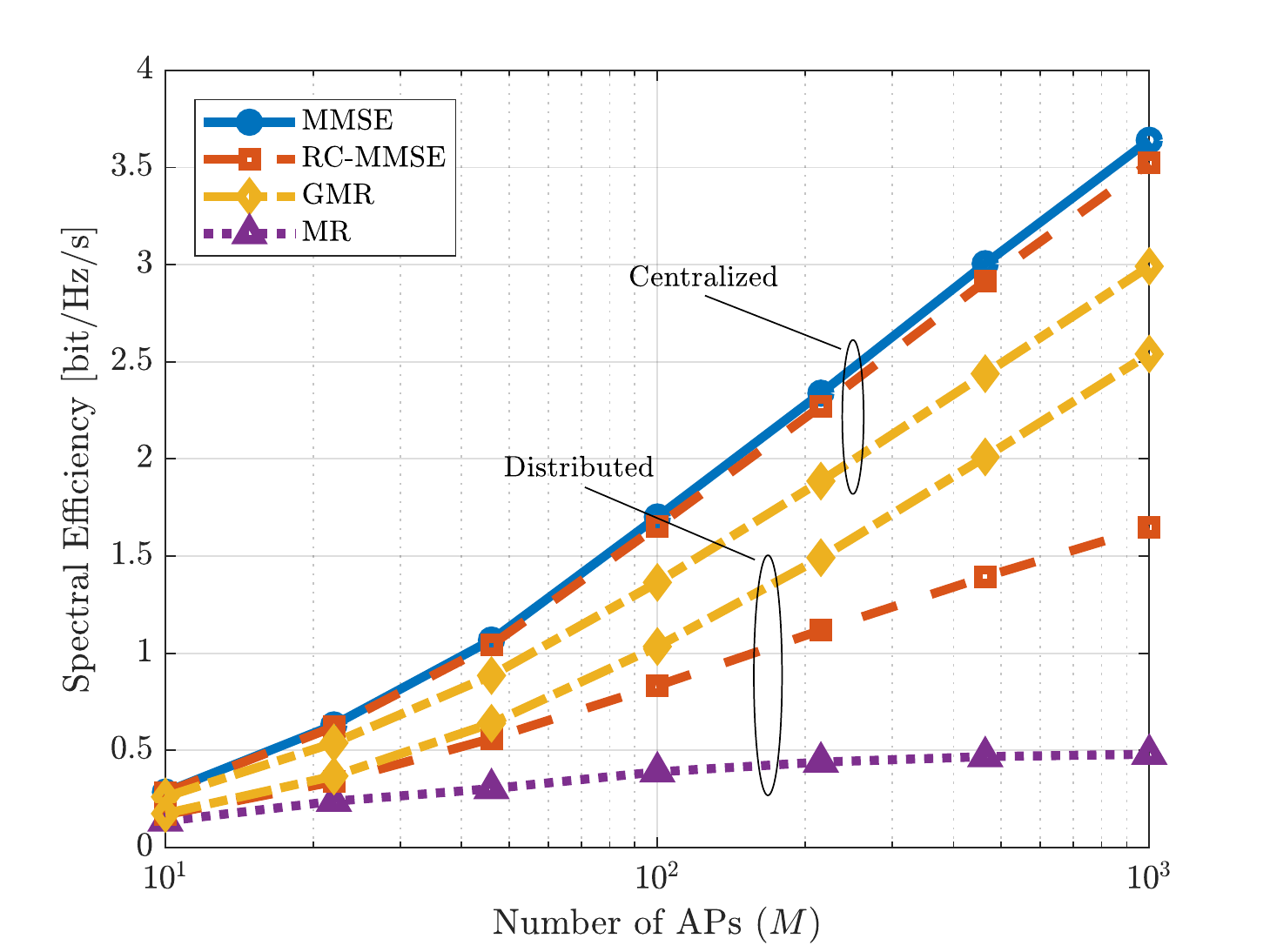}
	\caption{Average UL SE per UE for $K=5$.}\vspace{-0.6cm}
	\label{fig:M_range}
\end{figure}

Fig. \ref{fig:M_range} shows the asymptotic behavior of the average \gls{UL} \gls{SE} per \gls{UE} for $K=4$ with increasing $M$. The \glspl{UE} are located in the center of the area being $5$ m apart from each other with $D=1000$ m. The optimal \gls{MMSE} shows the best performance and confirms that \gls{CF} \gls{mMIMO} networks can indeed achieve unbounded capacity in the asymptotic regime (as stated in Proposition 1). Likewise, both \gls{RCMMSE} and \gls{GMR} centralized combiners \gls{SE} grows unboundedly with $M$. In agreement to Proposition 2, the distributed approach of \gls{GMR} shows the same asymptotic behavior, though the \gls{UatF} bound considerably underestimates its performance. Lastly, the poor \gls{SE} showed by the \gls{RCMMSE} proves that it is not an adequate scheme under statistical \gls{CSI}.
%\vspace{-0.5cm}
\section{Conclusions}\label{sec:conclusion}
This letter focused on pilot decontamination in \gls{CF} \gls{mMIMO}. Building on linear independence of correlation matrices, the \gls{RCMMSE} combining was proposed as an alternative to centralized \gls{MMSE} that almost matches the optimal performance and is asymptotically optimal. This scheme turns out to be a more than fitting approach for centralized computationally-limited networks. The \gls{GMR} alternative scheme was also derived showing the same asymptotic behavior but maximizing the \gls{UatF}. 
This approach is the upper-bound of \gls{MR}-based \gls{CF} networks. Both schemes are valid for a finite number of \glspl{AP} and have linear increasing complexity (in contrast to the optimal \gls{MMSE}). By the \gls{UL}-\gls{DL} duality, the same performance can be achieved in \gls{DL} where the \gls{GMR} alternative serves as an optimal approach for \gls{MR}-based precoders.
%\vspace{-0.4cm}
\section*{Appendix}
We define $\widehat{\bf H} = [\widehat{\bf h}_1,\ldots,\widehat{\bf h}_K]\in \mathbb{C}^{M\times K}$ and use the matrix inversion lemma to get (after multiplying and dividing by $M$)
\begin{align}\label{eq:SINR_maximizer_2}
	\check{\bf v}_k  \triangleq \frac{1}{M}{\bf Z}^{-1}\widehat{\bf H}\bigg(\frac{1}{M}\widehat{\bf H}^H {\bf Z}^{-1} \widehat{\bf H}+\frac{1}{M} {\bf I}_K\bigg)^{-1}\!\!\!{\bf e}_k
\end{align}
from which $\check{v}_{mk}$ is obtained as (by neglecting the factor $1/M$, since it does not affect the SINR in~\eqref{eq:SINR_CSI}) 
\begin{align}\label{eq:SINR_maximizer_2_1}
	\check{v}_{mk} \triangleq \frac{1}{{z}_m}\widehat{\bf h}_m^T\bigg(\frac{1}{M}\widehat{\bf H}^H {\bf Z}^{-1} \widehat{\bf H}+\frac{1}{M} {\bf I}_K\bigg)^{-1}\!\!\!{\bf e}_k
\end{align}
with $\widehat{\bf h}_m = [\widehat{h}_{m1},\ldots,\widehat{h}_{mK}]^T\in \mathbb{C}^{K}$. By exploiting the fact that, as $M\to \infty$,
\begin{align}\notag
	\frac{1}{M}\bigg[\widehat{\bf H}^H {\bf Z}^{-1} \widehat{\bf H}\bigg]_{\ell j} \!\!\!= \frac{1}{M}  {\widehat{\bf h}_{\ell}}^H{{{\bf Z}}^{-1}}\widehat{\bf h}_j \asymp  \frac{1}{M} {\rm tr}\left({\bf R}_j\bs\Psi^{-1}{\bf R}_{\ell}{{\bf Z}}^{-1}\right)
\end{align}
the combiner ${\bar {v}}_{mk}$ defined in~\eqref{eq:combiner_RCMMSE} follows.
% as follows:
%\begin{align}\label{eq:optimal_DMMSE_1}
%{\bar {v}}_{mk} =\frac{1}{{z}_m}\widehat{\bf h}_m^T{\boldsymbol{\varsigma}}_k = \frac{1}{{z}_m} \left(\sum_{i=1}^K\varsigma_{ki}\widehat{ h}_{mi}\right)\!
%\end{align}
%where $\varsigma_{ki} = \left[{\boldsymbol{\varsigma}}_k\right]_{i}$ and
%${\boldsymbol{\varsigma}}_k = \big({\bf B}+ {1}/{M}{\bf I}_K\big)^{-1}\!\!\!{\bf e}_k$
%with ${\bf B}\in \mathbb{C}^{K\times K}$ and $\big[{\bf B}\big]_{\ell j}\triangleq \beta_{j \ell}$. 

\bibliographystyle{IEEEtran}
\bibliography{asymptotic_cell_free_bib}

% Generated by IEEEtran.bst, version: 1.14 (2015/08/26)
\begin{thebibliography}{10}
\providecommand{\url}[1]{#1}
\csname url@samestyle\endcsname
\providecommand{\newblock}{\relax}
\providecommand{\bibinfo}[2]{#2}
\providecommand{\BIBentrySTDinterwordspacing}{\spaceskip=0pt\relax}
\providecommand{\BIBentryALTinterwordstretchfactor}{4}
\providecommand{\BIBentryALTinterwordspacing}{\spaceskip=\fontdimen2\font plus
\BIBentryALTinterwordstretchfactor\fontdimen3\font minus
  \fontdimen4\font\relax}
\providecommand{\BIBforeignlanguage}[2]{{%
\expandafter\ifx\csname l@#1\endcsname\relax
\typeout{** WARNING: IEEEtran.bst: No hyphenation pattern has been}%
\typeout{** loaded for the language `#1'. Using the pattern for}%
\typeout{** the default language instead.}%
\else
\language=\csname l@#1\endcsname
\fi
#2}}
\providecommand{\BIBdecl}{\relax}
\BIBdecl

\bibitem{Marzetta10}
T.~L. {Marzetta}, ``{Noncooperative Cellular Wireless with Unlimited Numbers of
  Base Station Antennas},'' \emph{IEEE Trans. Wireless Commun.}, vol.~9,
  no.~11, pp. 3590--3600, November 2010.

\bibitem{Bjornson18_unlimited}
E.~{Bj\"ornson}, J.~{Hoydis}, and L.~{Sanguinetti}, ``{Massive MIMO Has
  Unlimited Capacity},'' \emph{IEEE Trans. Wireless Commun.}, vol.~17, no.~1,
  pp. 574--590, 2018.

\bibitem{Ngo17}
H.~Q. Ngo, A.~Ashikhmin, H.~Yang, E.~G. Larsson, and T.~L. Marzetta,
  ``{Cell-Free Massive {MIMO} Versus Small Cells},'' \emph{IEEE Trans. Wireless
  Commun.}, vol.~16, no.~3, pp. 1834--1850, 2017.

\bibitem{Nayebi17}
E.~Nayebi, A.~Ashikhmin, T.~L. Marzetta, H.~Yang, and B.~D. Rao, ``{Precoding
  and Power Optimization in Cell-Free Massive {MIMO} Systems},'' \emph{IEEE
  Trans. Wireless Commun.}, vol.~16, no.~7, pp. 4445--4459, 2017.

\bibitem{Bjornson20}
E.~{Bj\"ornson} and L.~{Sanguinetti}, ``{Making Cell-Free Massive MIMO
  Competitive With MMSE Processing and Centralized Implementation},''
  \emph{IEEE Trans. Wireless Commun.}, vol.~19, no.~1, pp. 77--90, 2020.

\bibitem{Sanguinetti18}
L.~{Sanguinetti}, E.~{Bj\"ornson}, and J.~{Hoydis}, ``{Fundamental Asymptotic
  Behavior of (Two-User) Distributed Massive MIMO},'' in \emph{Proc. 2018 IEEE
  Global Commun. Conf. (GLOBECOM)}, 2018, pp. 1--6.

\bibitem{DBS21-book}
\BIBentryALTinterwordspacing
{\"O}.~T. Demir, E.~Bj{\"o}rnson, and L.~Sanguinetti, ``{Foundations of
  User-Centric Cell-Free Massive MIMO},'' \emph{Foundations and Trends in
  Signal Processing}, vol.~14, no. 3-4, pp. 162--472, 2021. [Online].
  Available: \url{http://dx.doi.org/10.1561/2000000109}
\BIBentrySTDinterwordspacing

\bibitem{Buzzi17}
S.~Buzzi and C.~D'Andrea, ``{Cell-Free Massive MIMO: User-Centric Approach},''
  \emph{IEEE Wireless Commun. Lett.}, vol.~6, no.~6, 2017.

\bibitem{Sanguinetti2020}
L.~Sanguinetti, E.~Bj{\"o}rnson, and J.~Hoydis, ``{Toward {Massive MIMO 2.0}:
  Understanding Spatial Correlation, Interference Suppression, and Pilot
  Contamination},'' \emph{IEEE Trans. Commun.}, vol.~68, no.~1, 2020.

\bibitem{Neumann18}
D.~Neumann, T.~Wiese, M.~Joham, and W.~Utschick, ``{A Bilinear Equalizer for
  Massive {MIMO} Systems},'' \emph{IEEE Trans. Signal Process.}, vol.~66,
  no.~14, pp. 3740--3751, 2018.

\end{thebibliography}

\end{document}